\documentclass[preprint2,times,tighten]{aastex6}

\usepackage{color}
\usepackage{enumitem}
\usepackage{hyperref}
\usepackage{verbatim}
\usepackage{amsmath,amstext,mathrsfs}
\usepackage[all]{hypcap} 
\usepackage{afterpage}   
\usepackage{marginnote}
\usepackage{makecell}
\usepackage{multirow}
\renewcommand{\arraystretch}{1.5} 
\newcommand*{\ch}{} % This is Chenghan's color
\def \coeight {C$^{18}$O}
\def \cotw {$^{12}$CO}
\def \coth {$^{13}$CO}
\def \s {\hspace{0.5cm}}

\begin{document}
\title{
%Tracing magnetic field morphology using the Velocity Gradient Technique in multiple self-gravitating molecular tracer maps
{\ch Tracing magnetic field morphology using the Velocity Gradient Technique in the presence of CO self-absorption}}
\author{Cheng-han Hsieh\altaffilmark{1}, Yue Hu\altaffilmark{2,3}, Shih-Ping Lai\altaffilmark{1,4}, Ka Ho Yuen\altaffilmark{2}, Sheng-Yuan Liu\altaffilmark{4}, I-Ta Hsieh\altaffilmark{4}, Ka Wai Ho\altaffilmark{5}, A. Lazarian\altaffilmark{2}}
\email{lazarian@astro.wisc.edu}
\altaffiltext{1}{Department of Physics, National Tsing Hua University, Taiwan}
\altaffiltext{2}{Department of Astronomy, University of Wisconsin-Madison, USA}
\altaffiltext{3}{Department of Physics, University of Wisconsin-Madison, USA}
\altaffiltext{4}{Institute of Astronomy of Astrophysics, Academia Sincia, Taiwan}
\altaffiltext{5}{Chinese University of Hong Kong, Hong Kong}

\begin{abstract}
Probing magnetic fields in self-gravitating molecular clouds are generally difficult even with the use of the polarimetry. Based on the properties of magneto-hydrodynamic (MHD) turbulence and turbulent reconnection, Velocity Gradient Technique (VGT) provides a new way in tracing magnetic field orientation and strength based on the spectroscopic data. Our study tests the applicability of VGT in various molecular tracers, e.g. $^{12}$CO, $^{13}$CO, and {\ch C}$^{18}$O. By inspecting synthetic molecular line maps of CO isotopologue generated through radiative transfer calculations, we show that the VGT method can be successfully applied in probing the magnetic field direction in the diffuse interstellar medium as well as in self-gravitating molecular clouds. 
\end{abstract} 

\keywords{ISM: structure --- ISM: turbulence---magnetohydrodynamics (MHD) --- methods: numerical}
\section{Introduction}

Magnetized turbulence are of great importance in many astrophysical process inside interstellar medium (ISM) especially in the stage of cloud formation and evolution \citep{1977ApJ...218..148M,2005A&A...433....1A, 2012ARAA...50..29}. However, the study of magnetic fields in the ISM is complicated, since the ISM has multiple phases where the level of ionization, temperature, density, and molecular abundances change drastically \citep{2003LNP...614.....F}. The most common techniques to study magnetic fields are stellar light polarization, emission from aligned grains \citep{2015ARA&A..53..501A} and molecular line splitting (Zeeman effect, \citealt{2010ApJ...725..466C}). By
measuring the polarization from stars \citep{2000AJ....119..923H}, one can get some insight into the morphology of the galactic magnetic field. However, as it is possible only to sample magnetic fields in the direction towards the stars with known distances, this way of magnetic field sampling is limited. Dust polarization measurements, on the other hand, determine the direction of the projected magnetic field. They can also be used within Davis-Chandrasekhar-Fermi technique to roughly estimate the {\it plane-of-sky averaged} magnetic field strength \citep{1951PhRv...81..890D,1953ApJ...118..113C,2008ApJ...679..537F}. Unfortunately the measurement of magnetic field through dust polarization is in general difficult, since the grain alignment efficiency drops significantly in the case of high optical depth, which limits the reliability to trace magnetic field in optically thick regions \citep{2015ARA&A..53..501A}. Line splitting, such as the Zeeman effect directly measures the intensity of the line-of-sight magnetic field \citep{2008ApJ...685..281C,2010ApJ...725..466C} without any assumptions. The Zeeman splitting is a small fraction of the line width, and only the Stokes spectra can be detected; these spectra reveal the sign and magnitude of the line-of-sight component of magnetic field \citep{2008ApJ...680..457T}\textbf{.}  However, the Zeeman measurements require very high sensitivity and long integration times. Thus in many cases only upper limits of magnetic field strength can be obtained.  

The Velocity Gradient Technique (VGT) \citep{2017ApJ...835...41G, YL17a,YL17b,2017arXiv170303035G, LY18a} is a new technique that can measure the direction and intensity of the magnetic fields only using spectroscopic observations. VGT has its foundations in the theory of MHD turbulence which states that the velocity motions of turbulent fluids are anisotropic and the direction of anisotropy is determined by the local direction of the magnetic field, i.e. fluid motions in the presence of MHD turbulence are eddy-like and the axis of rotation is aligned with the magnetic field surrounding the eddy \citep{1995ApJ...438..763G, LV99}. This property of turbulent motion follows from the theory of turbulent reconnection \citep{LV99}.\footnote{The theory of turbulent reconnection predicts the violation of the accepted concept of flux freezing in turbulent conducting fluids (see more in Eyink et al. 2011, 2013) which has important concequences for star formation (Lazarian et al. 2012).} This theory predicts that magnetic fields do not present an impediment for eddies that are aligned with the magnetic field surrounding them. Therefore most of the energy of the of the turbulent cascade is channeled into these type of eddies that trace magnetic field direction. It is essential for the VGT that the alignment of the eddies happens with the {\it local} magnetic field direction rather than the mean of magnetic field direction. Incidentally the original theory in \cite{1995ApJ...438..763G} does not make this distinction\footnote{Formally the Goldreich \& Sridhar theory is formulated in the mean field of reference where the Goldreich \& Sridhar cricial balance relations, which are the corner stone of the theory, are not valid.}. The fact that the local system of reference should be used is obvious from the turbulent reconnection theory \citep{LV99} and is proved reliably in numerical simulations \citep{2000ApJ...539..273C,2001ApJ...554.1175M,2002ApJ...564..291C}. Due to the fact that eddies trace the local direction of magnetic field the VGT captures the detailed magnetic field structure within the turbulent volume under study. 

VGT was first applied to HI data \citep{YL17a} and then extended to the cases when gravity \citep{YL17b} and self-absorptions \citep{2017arXiv170303035G} are important. In \citet{YL17b} they presented a new smoothing method to estimate the gradients that presents a more reliable estimation, in which the VGT has still been used to measure the magnetic field direction in HI and $^{13}$CO. In its original formulation (see \cite{2017ApJ...835...41G}) the VGT used centroids as the observationally available proxy of the velocity destribution. Later, in \cite{LY18a} it was proposed to trace magnetic fields using gradients of intensity within the thin channel maps, i.e. using the Velocity Channel Gradients (VChGs). This is possible as the theory of the non-linear mapping of motions from a turbulent volume to the Position-Position-Velocity space {\ch \citep{LP00}} predicts that thin channel maps represent well the velocity statistics. The VChGs approach shows higher {\ch accuracy} compared to the analysis based on centroids. It also increases the utility of VGT as the per-channel gradient analysis allows observers distinguish molecular clouds along the line of sight in the galactic plane as well as to obtain the three-dimensional magnetic field tomography of HI using the galactic rotation curve \citep{2018arXiv180510329G}.

In parallel with VGT we also use its derivative technique, namely, Intensity Gradients (IGs). The IG technique should be distinguished from Histograms of Relative Orientation (HRO) technique proposed in 
\cite{Soler2013}. Within the IG approach (see \cite{YL17b}) one uses the procedures developed within the VGT technique, e.g. block averaging procedure, in order to reliably determine the gradient directions that can be compared to the magnetic field direction on the point-wise basis. In HRO, on the contrary, no point-wise comparison is possible, but only the comparison in terms of the statistics of the orientations as a function of column density is available. In general, density fluctuations are less direct statistics of turbulence compared to velocities (see {\ch \cite{2003MNRAS.345..325C}}) and therefore we expect less accurate magnetic field tracing with IGs compared to that available with the VGT. The synergy of using of the IGs and velocity gradients is discussed elsewhere.   

Depending on the physical scale of the observations, self-gravity can strongly affect the dynamics of the gas in GMCs. GMCs are studied through molecular transitions, most commonly those of $^{12}$CO, $^{13}$CO, and C$^{18}$O. It is therefore important to understand the effects of self-gravity using molecular Position-Position-Velocity (PPV) data in the VGT.  This work then builds on the separate studies of self-gravity and molecular emission in PPV data cubes. In this work,  we apply VGT to analyze two conditions, with and without self-gravity. By processing the ideal MHD simulations with a new radiative transfer code SPARX, We produce the synthetic maps of $^{12}$CO, $^{13}$CO, and C$^{18}$O.

These tracers are most prevailing in the diffuse ISM, and the differences of optical depths between these tracers could offer information of magnetic field along the line of sight.

In what follows, in Section \ref{sec:theory}, we discuss how VGT perform and to be optimized in the case of self-absorbing self-gravitating molecular tracer maps. In Section \ref{sec:data}, we give a brief view of the MHD simulation and the radiative transfer calculation. In Section \ref{sec:4}, we present our results about VGT in the presence of self-absorbing media and self-gravity.  In Section \ref{sec:6} we discuss the influence of radiative transfer in VGT. In Section \ref{sec:7}, we give our conclusions.
\begin{table*}[t]
\centering
\begin{tabular}{c | c | c | c | c | c | c| c | c | c}
\hline \hline
Model & $M_S$ & $M_A$ & $\beta=2M_A^2/M_S^2$ & {\cotw, IGs} & {\coth, IGs} & {\coeight, IGs} & {\cotw, VCGs} & {\coth, VCGs} & {\coeight, VCGs}\\ \hline \hline
b11 & 0.41 & 0.04 & 0.02 	& 34\s30 & 36\s32 & 43\s30 & 33\s35 & 33\s38 &35\s38\\
b12 & 0.92 & 0.09 & 0.02 	& 15\s10 & 19\s11 & 25\s16 & 14\s16 & 16\s18 &18\s25\\
b13 & 1.95 & 0.18 & 0.02 	& 7\s5 	 & 10\s7  & 17\s11 & 7\s7	& 10\s8 &17\s12\\
b14 & 3.88 & 0.35 & 0.02 	& 10\s8	 & 10\s11 & 20\s17 & 10\s9	& 12\s12 &16\s16\\
b15 & 7.14 & 0.66 & 0.02 	& NA\s18 & NA\s19 & NA\s28 & NA\s20 & NA\s21 &NA\s27\\ \hline
b21 & 0.47 & 0.15 & 0.2178 	& 28\s29 & 30\s33 & 39\s35 & 22\s26 & 29\s29 &35\s30\\
b22 & 0.98 & 0.32 & 0.2178 	& 18\s15 & 30\s25 & 31\s33 & 19\s18 & 22\s24 &24\s29\\
b23 & 1.92 & 0.59 & 0.2178 	& 14\s14 & 19\s20 & 22\s29 & 14\s15 & 18\s21 &22\s20\\ \hline
b31 & 0.48 & 0.48 & 2 		& 33\s37 & 68\s61 & 66\s61 & 37\s38 & 36\s39 &37\s36\\
b32 & 0.93 & 0.94 & 2 		& 30\s29 & 42\s46 & 43\s41 & 31\s35 & 43\s36 &46\s34\\ \hline
b41 & 0.16 & 0.49 & 18.3654 & 55\s57 & 59\s57 & 57\s56 & 50\s45 & 53\s45 &51\s45\\
b42 & 0.34 & 1.11 & 18.3654 & 36\s40 & 53\s48 & 54\s52 & 41\s39 & 46\s50 &44\s51\\ \hline
b51 & 0.05 & 0.52 & 200 	& 59\s58 & 58\s56 & 58\s57 & 54\s54 & 54\s54 &53\s55\\
b52 & 0.10 & 1.08 & 200 	& 64\s69 & 64\s67 & 64\s67 & 47\s48 & 47\s48 &46\s49\\ \hline \hline
\end{tabular}
\caption{\label{tab:sim} Description of the MHD simulation cubes.  $M_s$ and $M_A$ are the instantaneous values at each the snapshots are taken. Each MHD cube contains three types of absorbing media $^{12}$CO, $^{13}$CO, and C$^{18}$O with emission line J=1-0.  Both ideal case without self-gravity and the case with the presence of self-gravity are considered.  The resolution of each cube is $480^3$. The right 6 columns show the relative angles for Intensity Gradients(IGs) and Velocity Centroid Gradients(VCGs) methods such that under 68.27\% confidence interval (1 $\sigma$),the VGT predictions is the same as simulated B field. Inside the tuple, the first values show results for self-gravity data and the second values for cases without self-gravity.}
\end{table*}
\section{The theoretical expectation of VGT under different molecular tracer maps}
\label{sec:theory}
\subsection{Velocity Gradient Technique}

The development of VGT is highly related to the recent establishment of MHD turbulence theory through numerical studies. The core theoretical consideration is derived from \citet{1995ApJ...438..763G} \&  \citet{LV99}. \citeauthor{1995ApJ...438..763G} (\citeyear{1995ApJ...438..763G}, hereafter GS95) predicted the anisotropy of MHD turbulence, and \citet{LV99} introduced the theory of turbulent reconnection. \cite{LV99} shows that magnetic field lines are allowed to rotate perpendicularly around each other due to fast turbulent reconnection. Similarly, in the framework of turbulence, the turbulent eddies are not constrained from rotating perpendicular to the direction of magnetic field. As a result, in random turbulence driving, the eddy motions perpendicular to magnetic field lines become more probable since magnetic tension force resists any other types of magnetic field motion.

Incidentally, it raises the consideration of the importance of local magnetic field in respect to the motions of Alfvenic turbulence. The concept of the local system of reference was confirmed in \citet{2000ApJ...539..273C}. Alfvenic eddies that are not constrained by magnetic tension create a Kolmogorov
cascade with velocities $v_l\sim l_\perp^{\frac{1}{3}}$
, where $l_\perp$ is measured with respect to the local direction of the magnetic field. It is evident that the eddies mixing magnetic field lines perpendicular to their direction should induce Alfvenic waves along the magnetic field. Hence, it is essential for VGT to trace the local magnetic field around turbulent eddies rather than the mean magnetic field.

It is {\ch a} well-established fact that the statistics of Alfvenic turbulence is anisotropic along the local magnetic field directions \citep{LV99,2002ApJ...564..291C}.  Similarly, the velocity gradients also show a distribution of directions in which the most probable orientation of gradients is perpendicular to the magnetic field direction \citep{YL17a}. Hence, the direction of the magnetic field can be obtained by rotating the most probable orientation of gradients by 90 degrees.

In the framework of VGT, three types of two-dimensional maps are frequently used to trace magnetic fields in a number of context: Intensity maps \textbf{I(x,y)}, velocity centroid maps \textbf{C(x,y)}, and velocity channel maps \textbf{Ch(x,y)}. These maps are produced by doing integral along the velocity axis of the PPV (Position-Position-Velocity) cube for all tracers. In this work, we use \textbf{I(x,y)} and \textbf{C(x,y)}:
\begin{equation}
C(x,y)=\frac{\int dv\ \rho(x,y,v)\cdot v }{I(x,y)}
\label{eq1}
\end{equation}
\begin{equation}
I(x,y)=\int dv\ \rho(x,y,v)
\label{eq2}
\end{equation}
where $\rho$ is the PPV gas density, and $v$ is the velocity component along the line of sight.

Then the gradient angle at pixel $(x_{i},y_{j})$ is defined as:
\begin{equation}
\bigtriangledown_{i,j}=tan^{-1}\left[\frac{f(x_{i},y_{j+1})-f(x_{i},y_{j})}{f(x_{i+1},y_{j})-f(x_{i},y_{j})}\right]
\end{equation}
$f(x,y)$ can be either \textbf{I(x,y)} or\textbf{ C(x,y)}. This will make up the pixelized gradient field  of a spectroscopic data.

In \cite{YL17a} they propose the recipe of {\bf sub-block averaging} to predict the direction of magnetic field through gradients in a statistical region of interest. When the statistical samples are sufficiently large (In \cite{YL17a} and later \cite{LY18a} they provided a criterion to determine the {\ch optimal} block size for a given number of gradient statistics.), the histogram of gradient orientations would show a Gaussian profile. Within a block we obtained the most probable orientation which is the peak of the Gaussian corresponding to the local direction of the magnetic field within the block. The VGT technique uses the sub-block averaging method and it is getting the results that are very different from those that can be obtained with the Histogram of Relative Orientations (HRO) technique  \citep{Soler2013,Soler2017a}. The latter technique uses gradients of intensity and it requires polarimetry data to define the direction of the magnetic field, while the VGT is polarization-independent and complementary way  way of finding the magnetic field direction.

\subsection{The effect of radiative transfer}

It has already been demonstrated that the prediction of magnetic field from VGT shows a good alignment with the presence of absorbing media $^{13}$CO $J=2-1$ \citep{2017arXiv170303035G}. Aside from $^{13}$CO , $^{12}$CO and C$^{18}$O are also common tracers of interstellar molecular flows when the number density of neutral hydrogen (HI) is between $10^{2}$\,cm$^{-3}$--$10^{4}$\,cm$^{-3}$, which is the common density for newly born self-gravitating molecular cloud \citep{2012ARAA...50..29}. The most important difference between the isotopologous of CO in tracing the molecular flow is their optical depths. That means one can use $^{12}$CO  to trace the flow of molecular gases in the diffuse surrounding region of a self-gravitating molecular cloud due to its weaker penetration power while using $^{13}$CO and C$^{18}$O  to estimate the accumulated contribution of gas motions in a thicker line-of-sight cloud. 

Due to the differences in optical depths of CO isotopologues, it is possible to use VGT to stack the three-dimensional {\it tomography} from surrounding layers to deeper core layers. However, it is difficult to explore the magnetic field morphology through VGT when strong self-gravity is present since \cite{YL17b} \& \cite{LY18a} suggest that the gradients of intensities and centroids are gradually rotating from $\perp {\bf B}$ to $\parallel {\bf B}$ when the {\it stage of collapse} increases. The separation of diffuse and dense media through molecular tracer maps with different optical map assists observers to study the velocity anisotropy and thus magnetic field structure of the molecular cloud from the outermost diffuse layer to the dense core layer.

\section{Method}
\label{sec:data}
\subsection{MHD data}
The numerical 3D MHD simulations are obtained from code ZEUS-MP/3D \citep{2006ApJS..165..188H}, with a single fluid, operator-split, staggered grid MHD Eulerian assumption. The data has been used in \citet{2018arXiv180202984L,2018arXiv180208772H} to set up a three-dimensional, uniform, isothermal turbulent medium.

Periodic boundary conditions, as well as solenoidal turbulence injections, are applied in the simulation for emulating a part of the interstellar cloud. We employ various Alfvenic Mach numbers $M_{A}=\frac{V_{L}}{V_{A}}$ and sonic Mach numbers in our simulation $M_{S}=\frac{V_{L}}{V_{S}}$, where $V_{L}$ is the injection velocity, while $V_{A}$ and $V_{S}$ are the Alfven and sonic velocities respectively (See Table \ref{tab:sim} for details.). We shall refer to the simulations Table \ref{tab:sim} by their model name. For instance, our figures will have the model name indicating which data cube was used to plot the figure. The simulations are named with respect to a variation of the ascending values of $\beta$. The ranges of $M_{S}$, $M_{A}$, $\beta$ are selected so that they cover different possible scenarios of astrophysical turbulence from very subsonic to supersonic cases. For each cube, We considering both cases with and without self-gravity in this work.

\subsection{The radiative transfer calculation}

\begin{figure}[htb]
\centering
\includegraphics[width=0.82\linewidth,height=2.42\linewidth]{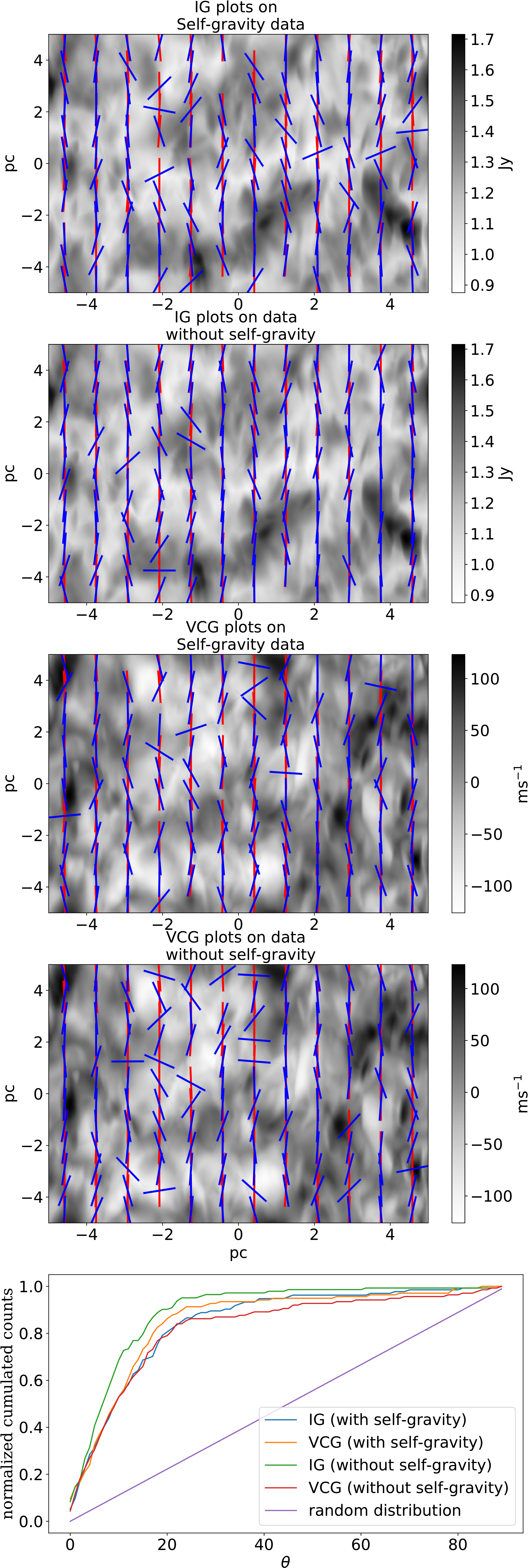}
\caption{$^{12}$CO b12 ($M_S=0.92$, $M_A=0.09$) data set. For each method (IGs and VCGs) and each data (with or without self-gravity) we plot the 2D vector plots and the statistical results. The blue vectors represent the projected  B fields from the simulation, and the red vectors represent the VGT predicted B field direction. {\ch The gray background represents the Moment 0 or Moment 1 of the data cube.} The relative angle between the simulated B field and the VGT predicted direction is shown in the normalized cumulative plots.}
\label{fig1}
\end{figure}

We performed three-dimensional LTE (local thermal equilibrium) radiative transfer on a Cartesian grid to generate synthetic maps with the SPARX (Simulation Package for Astronomical Radiative Xfer) code. The SPARX code is designed to calculate radiative transfer for both molecular line transitions and dust continuum and its details of the package are given in the Appdendix.

In the calculation, molecular gas density and velocity information are extracted from the MHD data mentioned in Section 3.1. A gas temperature of 10 K, which is typical in molecular clouds \citep{1997ApJ...483..210W}, is assumed. The fractional abundances of CO isotopologus $^{12}$CO,$^{13}$CO, and C$^{18}$O are set to be $1\times 10^{-4}$, $2\times 10^{-6}$, $1.7\times 10^{-7}$ respectively. The commonly used $^{12}$CO to H$_2$ ratio of $1\times 10^{-4}$ comes from the cosmic value of {\ch C\,/\,H} $= 3\times10^{-4}$ and the assumption of $15\%$ of C is in the molecular form. For the abundance of $^{13}$CO, we adopted a $^{13}$CO/$^{12}$CO ratio of $1/69$ \citep{1999RPPh...62..143W}. Hence, the $^{13}$CO to {\ch H}$_2$ ratio is approximated to $2\times 10^{-6}$. With $^{12}O/^{18}O = 500$ \citep{ToolsofRadio_astronomy}, we obtain C$^{18}$O to H$_2$ ratio to be $1.7\times 10^{-7}$.

When producing the synthetic molecular channel maps, we focus on the lowest transition $J=1-0$ of CO isotopologues, in which the LTE condition is satisfied. The required (critical) density for thermally populating the $J=1-0$ of CO isotopologues is $\sim$ 10$^3$ cm$^{-3}$, which is comparable to the molecular gas density in the diffuse ISM. The high optical depth of $^{12}$CO $J=1-0$ transition further facilitates the reduction of the required (critical density) for LTE population.

When applying LTE assumption to $^{13}$CO on cloud models a factor of 2 uncertainties on the column density derived should be expected \citep{Dishoeck1992}. For C$^{18}$O, the molecular line is optically thiner than the other two species allowing the tracer to trace into the denser regions of the clouds.

\begin{figure}[htb]
\centering
\includegraphics[width=.82\linewidth,height=2.42\linewidth]{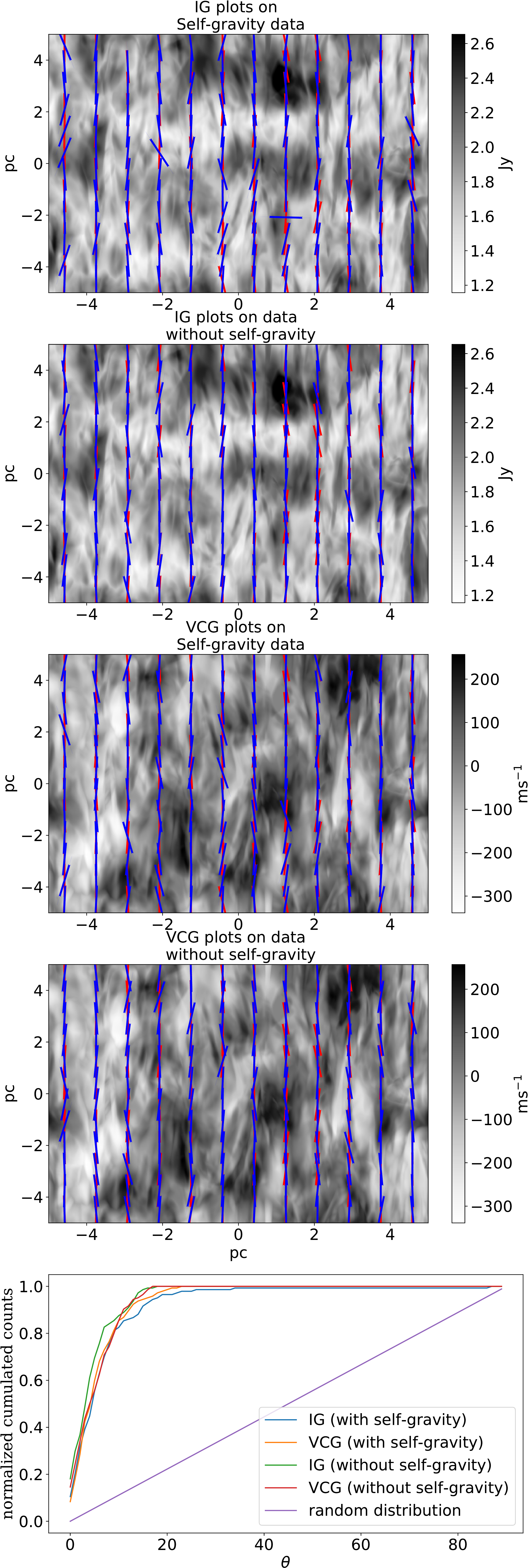}
\caption{$^{12}$CO b13 ($M_S=1.95$, $M_A=0.18$) data set. For each method (IGs and VCGs) and each data (with or without self-gravity) we plot the 2D vector plots and the statistical results. The blue vectors represent the projected B fields from the simulation, and the red vectors represent the VGT predicted B field direction.{\ch The gray background represents the Moment 0 or Moment 1 of the data cube.} The relative angle between the simulated B field and the VGT predicted direction is shown in the normalized cumulative plots.}
\label{fig2}
\end{figure}

\section{Responses from different optical tracers} 
\label{sec:4}

\begin{figure}[htb!]
\centering
\includegraphics[width=0.82\linewidth,height=2.42\linewidth]{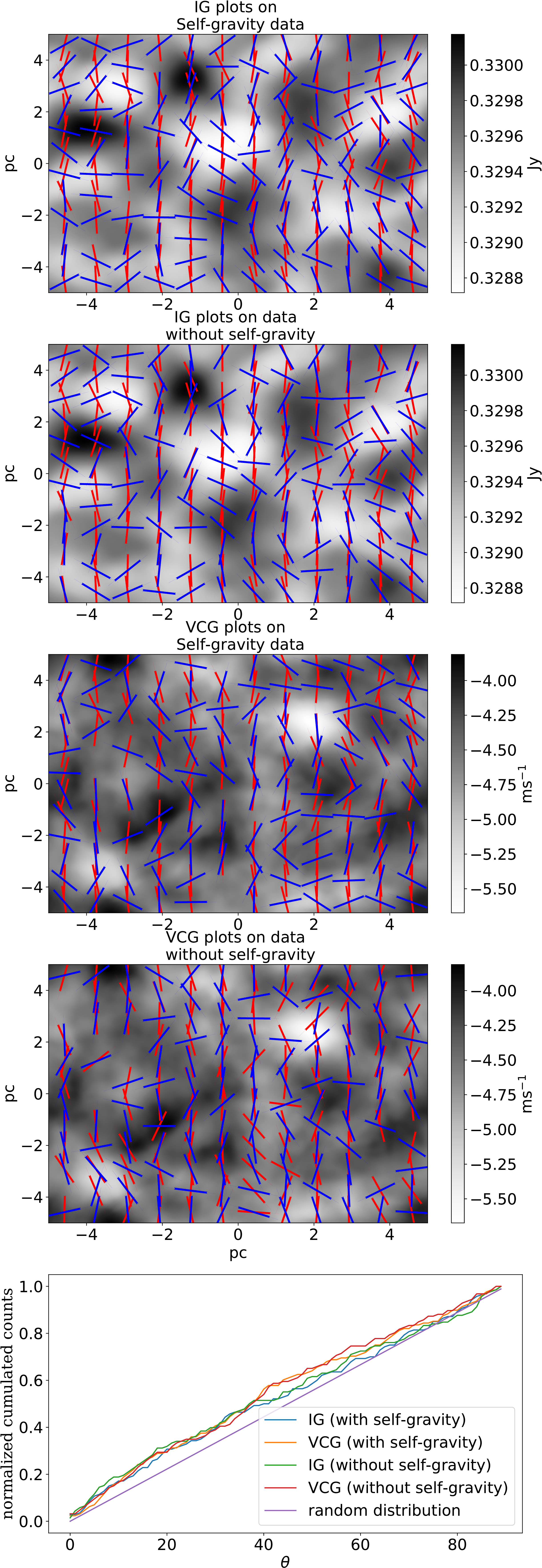}
\caption{$^{12}$CO b51 ($M_S=0.05$, $M_A=0.52$) data set. For each method (IGs and VCGs) and each data (with or without self-gravity) we plot the 2D vector plots and the statistical results. The blue vectors represent the projected B fields from the simulation, and the red vectors represent the VGT predicted B field direction. {\ch The gray background represents the Moment 0 or Moment 1 of the data cube.} The relative angle between the simulated B field and the VGT predicted direction is shown in the normalized cumulative plots.}
\label{fig3}
\end{figure}

\subsection{Effects of self-gravity}
\label{sec:4.1}

After carrying out radiative transfer simulations for each model with SPARX, we produce different position-position-velocity (PPV) data cubes for different tracers. We then compute integrated intensity map (moment 0 map) and velocity centroid map (moment 1 map) for each PPV cube (See \S \ref{sec:data} for more data details). After obtaining the moment maps, we then apply the VGT recipe \citep{YL17a,YL17b,LY18a} to obtain the prediction of B-field orientations. The recipe consists of the gradient operators, sub-block averaging \citep{YL17a} and also the error estimation method \citep{LY18a}. Inside each box, we collect all the directions predicted by the gradients and use a shifted Gaussian function \citep{2018arXiv180202984L} to fit. In principle, the Gaussian profile with a shifting term is a better representation on the gradient orientation distribution in numerical simulations, since the {\ch angle independent shifting term} is usually much higher than zero. For instance, with $M_A \sim 0.2$ the velocity iso-contour axis-ratio can be in hundreds \citep{2018arXiv180200987X}. However, simulations nowadays have limited resolution. There is a natural tendency for velocity contours to have smaller axes ratio due to the unresolvable minor axis. As a result, the count that away from the peak of gradient orientation distribution would not be close to zero. A constant shift would address the issue of finite length. In practice, the shift will not change the prediction of peak location by block-averaging. We set {\ch a} criteria on such that if the random area (shifted region) is greater than the Gaussian area then that block will not give predictions. We select a box size of $40^2$ pixels for the Gaussian fitting process, meaning that the sub-block averaged direction is given by the peak of the Gaussian fitting function on the gradient orientation histograms obtained from the $40 \times 40$ block.

\begin{figure}[tbh!]
\centering
\includegraphics[width=0.82\linewidth,height=2.42\linewidth]{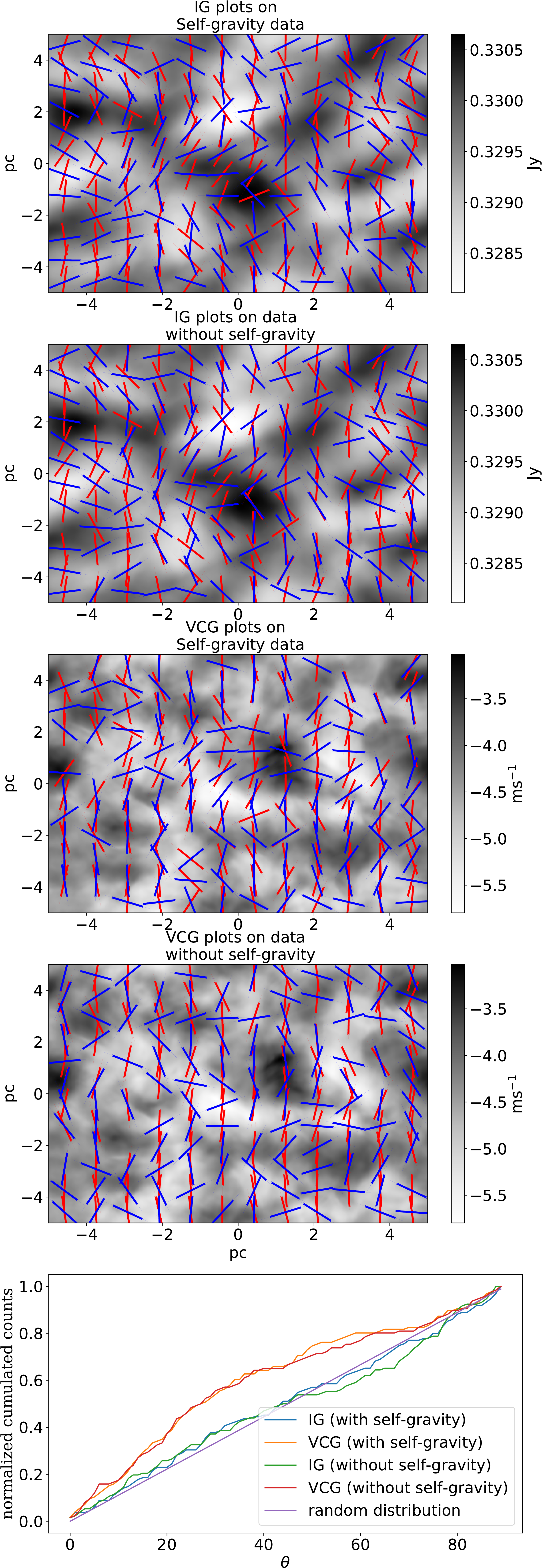}
\caption{$^{12}$CO b52 ($M_S=0.10$, $M_A=1.08$) data set. For each method (IGs and VCGs) and each data (with or without self-gravity) we plot the 2D vector plots and the statistical results. The blue vectors represent the projected B fields from the simulation, and the red vectors represent the VGT predicted B field direction. {\ch The gray background represents the Moment 0 or Moment 1 of the data cube.} The relative angle between the simulated B field and the VGT predicted direction is shown in the normalized cumulative plots.}
\label{fig4}
\end{figure}

The average value from the Gaussian fitting will be used to represent the VGT predictions in that box. The block-averaged vectors are then rotated 90 degrees to indicate the direction of magnetic field as predicted by VGT. We plot the results for model b12, b13, b51, and b52 in \autoref{fig1}, \autoref{fig2}, \autoref{fig3}, and  \autoref{fig4} respectively. In the simulations, we have three-dimensional data on magnetic fields. By mimicking dust polarization, we calculated the density weighted averaged B fields in the plane of sky and then compare these with VGT predictions to obtain relative angle. To allow better comparison, we collect all the relative angles between VGT predictions and simulated magnetic fields and present the results as cumulative plots. 

\autoref{fig1} and \autoref{fig2} are two examples showing that VGT method works very well. IGs means the gradient is computed from a moment 0 map and VCGs means that the gradient is calculated from moment 1 map. In \autoref{fig2}, we can see that the normalized cumulative counts quickly increases to above 0.9 within 20 degrees relative angle. This means 90\% of the VGT predicted B field vectors in these models have relative angles less than 20 degrees. As for b51 and b52 in \autoref{fig3} and \autoref{fig4}, the normalized cumulative counts are close to a random distribution. {\ch As later shown in \autoref{fig5}, these models fall close to the random distribution line.} A random distribution of the cumulative relative angle orientation histogram would indicate that VGT fails to provide a reliable prediction of magnetic field direction in the region of interest. 

In order to compare the performance of VGT across models, we define a parameter to determine whether the predictions are reliable. From each cumulative plot, we found the relative angle between VGT predictions and simulated B fields such that 68.27 percentage (1 $\sigma$) of vectors are within this relative angle range. The relative angles found under this criteria represent the uncertainty of the VGT method under 1 $\sigma$ confidence interval. We then plot the results for each model and each tracer in \autoref{fig5} and \autoref{fig6}.

\autoref{fig5} represents the response of the VGT method for simulation with self-gravity data. The blue circle represents the \cotw, the orange square represents the $^{13}$CO and the green triangle represents the \coeight. {\ch The dashed horizontal line represents the random distribution.} The upper panels represent the IGs and the lower panels represent the VCGs. \autoref{fig6} represents the result of simulations without self-gravity. 

Comparing \autoref{fig5} and \ref{fig6}, we found that the gradient vectors in the case of self-gravity are less aligned to the magnetic field direction compared to the case without self-gravity. Quantitatively most of the differences between the two cases are within 4$^{\circ}$.

  The difference between self and without self-gravity is negligible. This indicates that in the diffuse region VGT method can still be applied very well in the self-gravitating molecular clouds. The density tested from all 14 models have ranges between $0.003 \sim 50$ solar mass per $pc^3$, the corresponding hydrogen number density is $ 0.004 \sim 70$ cm$^{-3}$. To put this into observational perspectives, the giant molecular clouds have density $n(H_2)(cm^{-3})$ of 100, molecular clouds have the density of 300, molecular clumps have the density around $10^{3}$, and cloud cores have density $10^{5}$ \citep{Bodenheimer2011}. Thus the tested simulation cubes can be applied to giant molecular clouds which have similar number densities.

%%Note: 50 solar mass = 9.9x10**33kg; 1pc^3= 2.9*10^55cm^3; 1 mole = 10^23; --> ~72/cm^3

\subsection{Response of different tracers}
\label{sec:4.2}
% KH: Parameter setting

\begin{figure*}[htb]
\centering
\includegraphics[width=1.0\linewidth,height=.55\linewidth]{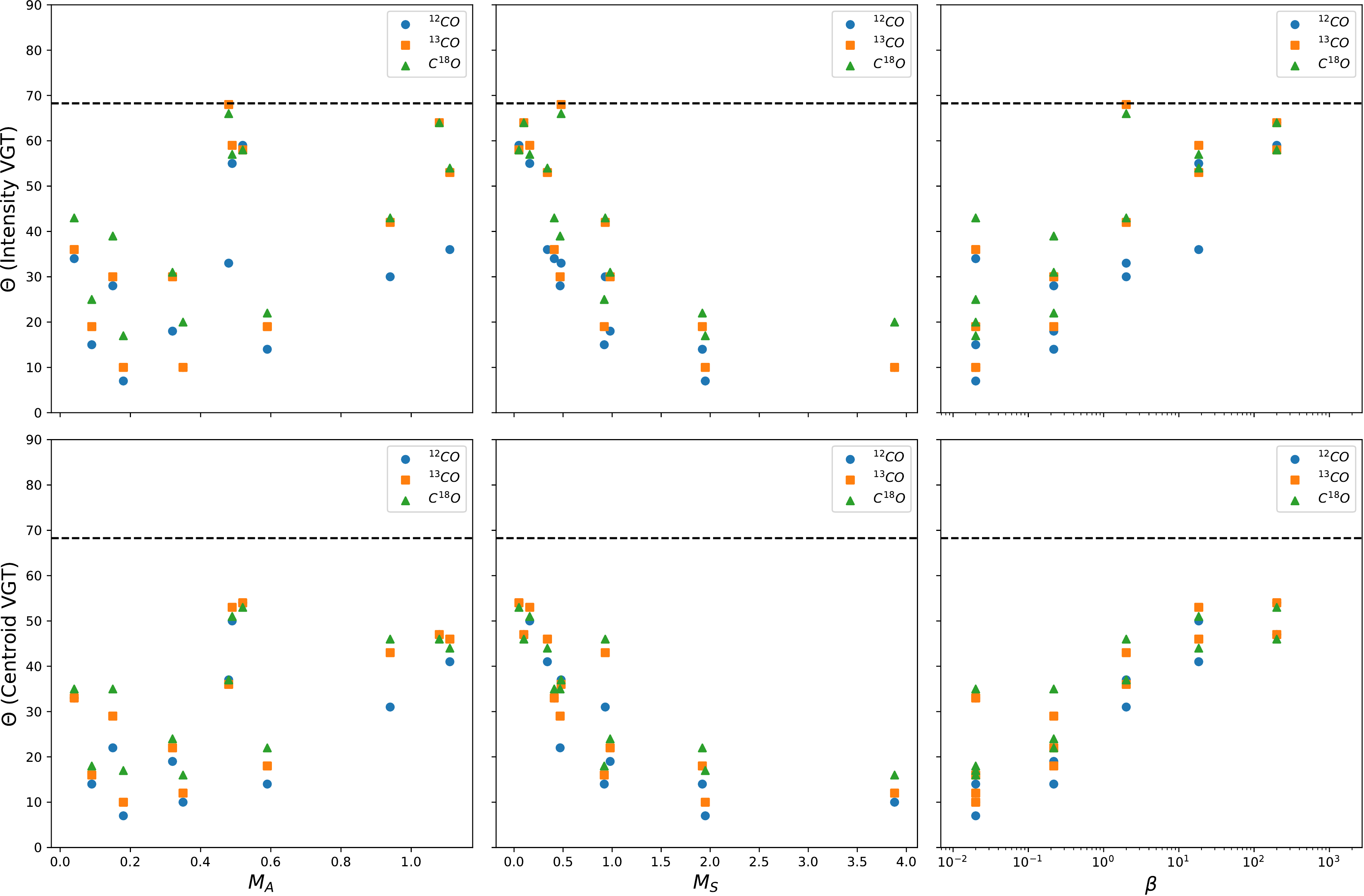}
\caption{The plot of relative angle (between rotated gradients and magnetic {\ch field}) such that the accuracy of IGs (top row) and VCGs (bottom row) is under 1 $\sigma$ uncertainty, with respect to the variation of Alfven Mach number ($M_A$, left column), Sonic Mach number ($M_S$, middle column), and compressibility $\beta=2(M_A)^2/(M_S)^2$ (right column). {\ch Random distribution is marked out as horizontal dash lines.} Turbulent MHD simulation data with self-gravity is used. 
}
\label{fig5}
\end{figure*}

\begin{figure*}[htb]
\centering
\includegraphics[width=1.0\linewidth,height=.55\linewidth]{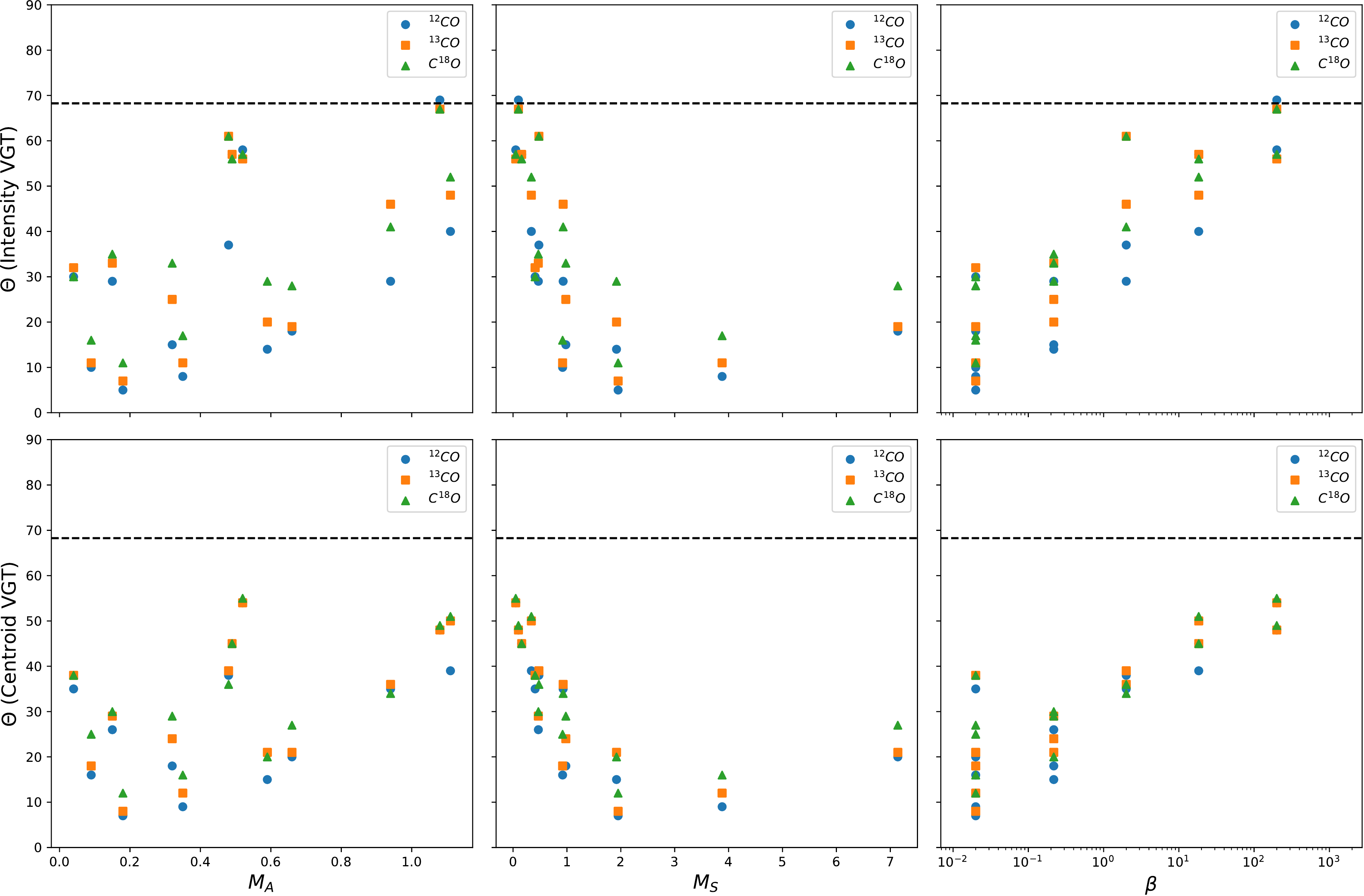}
\caption{The plot of relative angle (between rotated gradients and magnetic {\ch field}) such that the accuracy of IGs (top row) and VCGs (bottom row) is under 1 $\sigma$ uncertainty, with respect to the variation of Alfven Mach number ($M_A$, left column), Sonic Mach number ($M_S$, middle column), and compressibility $\beta=2(M_A)^2/(M_S)^2$ (right column). {\ch Random distribution is marked out as horizontal dash lines.} Turbulent MHD simulation data without self-gravity is used. 
}
\label{fig6}
\end{figure*}

In \autoref{fig5} and \autoref{fig6}, we plot the relative angle of each model such that the accuracy of VGT is under 1 $\sigma$ uncertainty, with respect to the variation of Alfven Mach number ($M_A$), Sonic Mach number ($M_S$), and compressibility $\beta=2(M_A)^2/(M_S)^2$. In the case of $\beta$ and $M_S$ respectively, we can find a strong linear correlation between $\beta$ or $M_S$ and the accuracy. With either the increasing of $M_S$ or the decreasing of $\beta$, VGT gives more accurate results. As for $M_A$, the overall plot is more scattered so the correlation is not clear. Since $\beta$ is proportional to $M_A$ but inversely proportional to $M_S$, the increasing of $M_S$ leads to a smaller $\beta$. Hence, $M_S$ is the dominating factor which leads to a more accurate tracing. In the central panels in \autoref{fig5} and \autoref{fig6}, as the $M_S$ increases to 2, the VGT predictions becomes better (smaller uncertainty). However, further increases of $M_S$ after 2 will slightly increase uncertainty. This is the same for both IGs and VCGs in with or without self-gravity conditions. 


\begin{figure}[h]
\centering
\includegraphics[width=1.0\linewidth,height=.7\linewidth]{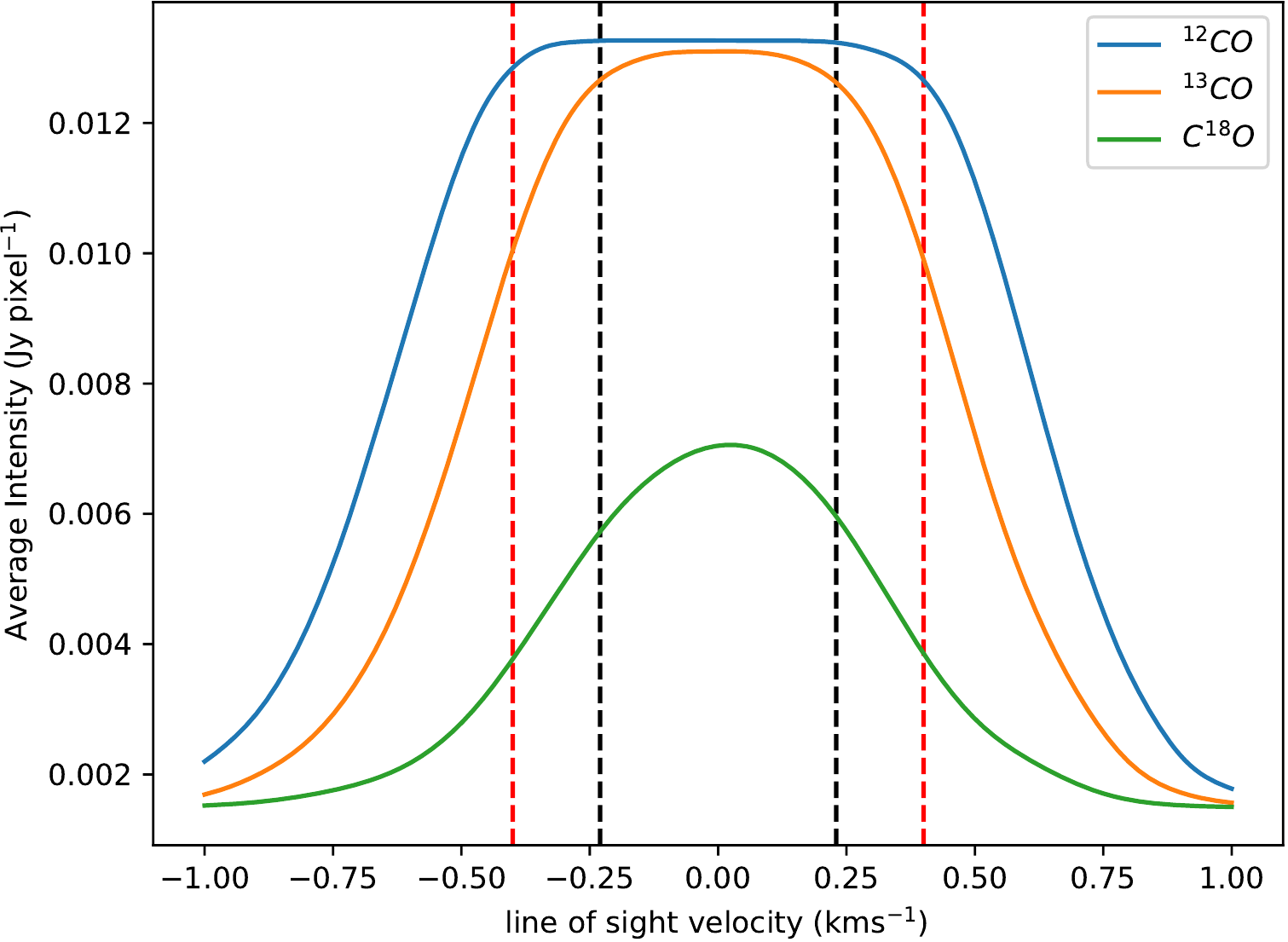}
\caption{Spectrum of b13 with self-gravity model after radiative transfer calculation. The blue line represents \cotw, the orange line represents the \coth, and the green line represents the \coeight. The black and red dash lines marked the region that is optically thick and the intensity is saturated for $^{13}$CO and $^{12}$CO respectively.
}
\label{fig7}
\end{figure}

In the right panels of \autoref{fig5} and \autoref{fig6}, a large scattering is observed when the $\beta$ value is small ($\sim 10^{-2}$). As for the $M_A$ responses, the overall correlation is weaker than $\beta$ and $M_S$. Before $M_A=0.2$, the VGT predictions, in general, have higher accuracy as $M_A$ increases. After $M_A=0.2$, the accuracy of the VGT method drops as $M_A$ increases further. In both plots, there are concave down relationships with minimum points (at $M_A=0.2$ and $M_S=2$). However, there is a sharp increase in uncertainty for $M_{A}=0.5$ but for $M_{S}$ plots there is no obvious feature similar to $M_{A}$. %{\yh yh: what about $M_S$=1.0?}{\ch I will say that one is not certain to claim, as the points are very scattered.}.

For observational purposes, we set a criterion such that under 1 $\sigma$ (68.27\%) confidence interval, the VGT predictions are within $20^{\circ}$ angle deviation from the true simulated magnetic field. Above this criteria, the VGT can give very accurate magnetic field predictions. Under this criteria, in \autoref{tab:sim} we found that for $^{12}$CO and $^{13}$CO, the IGs is very accurate when $0.1 < M_A < 0.7$ (best when $M_A = 0.18$), $0.9 < M_S < 7.1$ (best when $M_S = 2$), and $0.02 < \beta < 0.2$. As for \coeight, IGs is very accurate when $0.2 < M_A < 0.4$ (best when $M_A = 0.18$), $2.0 < M_S < 3.9$ (best when $M_S = 2$), and $0.02 < \beta < 0.2$. For \coeight, the Sonic Mach number range $M_S$ for accurate IGs predictions is smaller than the others. This result is consistent to \autoref{tab:sim}, \autoref{fig5}, and \autoref{fig6}, which show that C$^{18}$O has a larger uncertainty compare to the other two tracers. As for the VCGs, the accurate tracing ranges of $M_S$ and $M_A$ are the same as IGs for \cotw, and \coeight. However, for $^{13}$CO the ranges are $0.9 < M_S < 3.9$, and $0.1 < M_A < 0.4$ which are smaller than $^{12}$CO and $C^{18}O$. The results are organized in \autoref{tab:result}.

\begin{table*}[tp]
\centering
\label{tab:performance_comparison}
\begin{tabular}{|c|c|c|c|c|c|c|}
\hline
\multirow{2}{*}{Method}&
\multicolumn{3}{c|}{Intensity Gradients}&\multicolumn{3}{c|}{ Velocity Centroid Gradients}\cr\cline{2-7}
&$^{12}CO$&$^{13}CO$&$C^{18}O$&$^{12}CO$&$^{13}CO$&$C^{18}O$\cr
\hline\hline
$M_A$ & [0.1, 0.7] & [0.1, 0.7] & [0.2, 0.4] & [0.1, 0.7] & [0.1, 0.4] & [0.2, 0.4]\cr\hline
$M_S$ & [0.9, 7.1] & [0.9, 7.1] & [2.0, 3.9]  & [0.9, 7.1] & [0.9, 3.9] & [2.0, 3.9]\cr\hline
\end{tabular}
\caption{\label{tab:result}The range of $M_A$ and $M_S$ such that under 1 $\sigma$ uncertainty, the relative angle between VGT method and true simulated B fields is less than 20$^{\circ}$.}
\end{table*}

\subsection{Velocity Centroid Gradients vs Intensity Gradients}
\label{sec:4.3}

In the previous sections, we studied how the VGT method changes with respect to Alfven Mach number $M_A$, Sonic Mach number $M_S$, and $\beta$ in the case of optically thick and thin media. In this section, we will inspect how the change of moment maps will alter the prediction of magnetic field using VGT. Fig. \ref{fig5} and \ref{fig6} show the mean relative angle deviation between the magnetic field predictions from both VGT and emulated dust polarizations for both self-gravitating and non self-gravitating simulations respectively. The results from the intensity gradients are shown in the upper row of Fig. \ref{fig5} and \ref{fig6}, while that for centroid gradients are shown in the lower row of both figures.

In general, the centroid gradients have slightly better performance when compared to the IGs. For $^{13}$CO's case, IGs and VCGs have very similar results. For model b23 ($M_S$=1.92, $M_A$=0.59) the VCGs has relative angle of 21$^{\circ}$ while the IGs gives uncertainty of 20$^{\circ}$ for the self-gravity data set. Under the 20$^{\circ}$ degree criteria, this resulted a difference of range shown in \autoref{tab:result}.

The VCGs shows a slightly better performance in C$^{18}$O {\ch than the} $^{13}$CO and \cotw.  The crucial reason behind is the velocity channels in the wing side of the spectral line is more velocity-like and diffuse \citep{LP00} when the channel width is optically thin. Since the wing side has more weighting compared to the central part of the spectral line in the process of computing the velocity centroid, the velocity centroid displays more velocity eddies information compared to the total intensity map. As a result, the VCGs is more accurate since velocity eddies are direct probe of turbulence data while that for density eddies are indirect consequences of fluid compressions \citep{2003MNRAS.345..325C}. 

In the diffuse region it is expected that gravity takes negligible role in regulating the gas motions in molecular clouds. As a result, the gas motions are channeled by the local magnetic field directions and therefore the assumption for VGT holds for diffuse media (See \S \ref{sec:theory}). However it is discussed in both \cite{YL17b} \& \cite{LY18a} that the gradient orientation changes from $\perp B$ to $\parallel B$ according to the {\it stage of collapse} (aka re-rotation). However, for the two sets of figures we shown in Fig. \ref{fig5} and \ref{fig6}, we spot negligible differences in terms of the mean angle deviations. Therefore we conclude that the effect of gravity in the self-gravity simulations used in Fig. \ref{fig5} is not strong enough to trigger the re-rotation of gradients.

\subsection{Saturation after radiative transfer and the possibility of 3D magnetic field mapping through different tracers}

We plot the velocity spectral {\ch profile} for three different tracers in \autoref{fig7}. After carrying out the radiative transfer calculation, we observed many interesting features. First of all, the intensity of $^{12}$CO is much higher than that of $^{13}$CO and also that of C$^{18}$O. This is expected in observation as the abundance of the CO isotopes are much lower than that of \cotw.

{\ch While the optical depth of $^{12}$CO is significantly higher than that of $^{13}$CO, it is possible for the gradients of $^{13}$CO tracer maps to perform better than that of $^{12}$CO in probing the direction of magnetic field around self gravitating core. This is because around the center of gravity of an observed molecular cloud, the number of pixels that are {\ch truthfully} sampled by $^{13}$CO are much higher than that of $^{12}$CO. Since the accuracy of gradients relies on the structures displayed by neighboring pixel values, therefore it is natural for the gradients of $^{13}$CO maps to {\ch perform} better in tracing magnetic fields than that of $^{12}$CO due to the differences of sampling size.}
% even though the optical depth for $^{12}$CO is significantly larger than that of $^{13}$CO

{\ch However, in the synthetic image cubes, the molecular intensity will be saturated in the line core where the relative velocity is small. This can be understood by the fact that when the medium is optically thick , we can't see through the medium and what we will observe will be the gases on the surface. In Fig. \ref{fig7} we use the vertical black dash line to roughly marked out the region that is optically thick. For \cotw, the medium is optically thick when  $v \in [-0.4\sim 0.4$ kms$^{-1}]$. As for \coth, the opaque region is between $-0.23\sim 0.23$ kms$^{-1}$ and there is no saturation in C$^{18}$O in our sample synthetic data in Fig. \ref{fig7}. The situation mentioned above changes in terms of synthetic maps as those improperly sampled pixels are now displayed as a constant (See Fig. \ref{fig7}). Those saturated pixels, while not related to the kinematic properties of molecular gas, form anisotropic structures. As a result, the appearance of these constant pixels {\it does not } decreases the tracing power for optically thick tracers and the alignment measure of optically thick tracers should be higher than that of optically thin tracers unless the density structures are strongly distorted by gravity.}

In the case of saturation, the contribution of those saturated channels (e.g. velocity channels with $v \in [-0.4\sim 0.4$ kms$^{-1}]$ in the case of \cotw) will be zero when performing gradients since they are just intensity maps having constant pixel values.  As a result, when computing the gradients of integrated intensity or velocity centroid maps, only the velocity channels outside the saturated regions will contribute. In theory, by using different tracers, we can trace magnetic fields in regions corresponding to a different line of sight velocities. However, the gradients of the material distribution in the velocity space is very difficult to utilize in building up the 3D structure of magnetic field since the effect of velocity caustics is taking effect \citep{LP00}. It would be convenient to make use of the fact that, when the molecular cloud is optical thick for a certain tracer, only the contribution with line-of-sight deepness $z<\tau$ will be positively contributing when using the VGT. As as result, VGT can only trace the B field associated with the surrounding gases in the molecular clouds. Therefore, a line-of-sight distance dependent magnetic field tomography can be achieved by stacking the VGT results from different molecular tracers. 

\section{Discussion}
\label{sec:6}

\subsection{Strength\textbf{s} and limitations of VGT in the self-absorbing media}

The Velocity Gradient Technique shows a nice adaptivity on numerous physical conditions, which provides a robust way to get magnetic field orientations with high accuracy.  From our study, VGT even performs excellently in the case of self-gravitating and self-absorbing media. However, understanding the limitations of VGT to work in self-absorbing media is crucial for the community to utilize and further develop the VGT with molecular tracer maps.

As we know, polarization data is one of the most reliable ways to obtain the magnetic field orientation. However, it is only universally available by some state-of-the-art interferometric instruments, which require prior knowledge of the dust grain alignments and usually also the understanding of background emission, and even the cost of obtaining such measurement in the ground or even in space. Fortunately, VGT makes uses of readily available spectroscopic data, such as CHIPASS synchrotron survey, HI4PI neutral hydrogen atom distribution survey and COMPLETE survey, to provide nearly equally accurate measurements of the magnetic field orientations, which can also be cross-checked by using different tracers on the same piece of observation data.  Previous series of papers \citep{2018arXiv180402732Y,YL17a,LY18a} already show that the VGT can provide comparable or even better field-tracing ability in both numerical and diffuse observational data. It is no doubt that VGT can be synergistically used with dust/synchrotron polarization data in determining the morphology of {\it projected} magnetic field \citep{Letal17,YL18}.

% The manuscript lacks acknowledgment of the restrictions of the VGT, as well as counterpoints to it. Although there is a section titled "Strength (sic) and limitations...", it mainly focuses on how the present study may be extended and what other techniques may be used to estimate the magnetic field direction, instead of discussing what could go wrong with the VGT.
%there are several limitations for the technique to apply.
{\ch While the VGT is very powerful in predicting the magnetic field morphology using just spectroscopic data, there remains limited for it to be applied. The first and most important issue for VGT is that the technique is based on anisotropic properties of turbulent MHD, so it can only be used for turbulent systems. From this work, we found out it works best for supersonic turbulence systems with $0.02 \leq \beta \leq 0.2$. For molecular clouds with density $n(H_2) \sim 100\,cm^{-3}$, the typical magnetic field strength is on the order of $\sim  10 \, \mu G$ \citep{2010ApJ...725..466C}. With a typical gas temperature of $\sim10$\, K \citep{1997ApJ...483..210W}, the turbulence sound speed is in the order $\sim 0.2$\, kms$^{-1}$. From these observational constrains, it can be shown the criteria $0.02 \leq \beta \leq 0.2$ are commonly satisfied in GMC.}

{\ch For observational implementations of VGT, it is crucial to take note that VGT is technique based on group statistics, i.e. the technique highly depends on the quality of the data and it can degrade the spatial resolution of B field predicted for at least 20 times (Lazarian \& Yuen (2018). Block-average technique with Gaussian fitting of gradient angles ensures the accuracy of VGT predictions on magnetic fields. A block size of 20 used in VGT requires $20^2$ of independent measurements, which corresponds to an area of $20^2$ beam size in observational term. The final inferred magnetic field direction represents the averaged magnetic field direction within the block. Thus, observational map size to beam size ratio constrains the number of magnetic field vectors that can be predicted by VGT. As for the selection of block size, it is determined by how many independent measurements (beam size) is needed in order for the gradient angle to reach a Gaussian-like distribution. Observers applying VGT should include enough independent measurements to ensure the shifted Gaussian distribution for block-averaging. This is a necessary condition for applying VGT on observational data. In addition, if the observation is very noise dependent (have strong correlated noise), the gradient technique cannot extract the magnetic field orientation even with the use of very large block size. Fortunately the shape of the gradient angle distribution will tell whether the statistics of gradients are qualified for deriving a meaningful prediction of magnetic field in the region of interest. }

{\ch The VGT also} faces limitations when dealing with optically thick molecular tracer maps since VGT is technically an edge-detection algorithm relying on the turbulence statistics theory\citep{1995ApJ...438..763G,LV99} in predicting the direction of magnetic field.  If the pixels in the molecular tracer map are saturated in the central part of the velocity channels such as the case of \cotw, then the ability of using gradients in these velocity channels will be significantly limited. The detailed theoretical and numerical study on saturation effects will be addressed in the future paper (C.-h. Hsieh et. al in prep). Comparatively, the polarization fraction in the case of optically thick clouds are usually low \citep{2010SPIE.7741E..0EF,2016ApJ...824..134F} since these clouds with low optical depth are usually self-gravitating. The prediction from polarization might not be helpful in determining the direction of magnetic field without referring to other independent measures.

{\ch  Gravitational collapsing is one of the main forces affecting the performance of VGT tracing the magnetic field in molecular clouds. In collapsing regions, the gradient field becomes distorted and the direction of gradients no longer aligns perpendicularly to the magnetic field, but parallel. To account for this, the compensatory re-rotation of gradients (i.e. rotating the gradient by 90$^{o}$ again) should be applied to the gradient \citep{survey}. 
}{\ch In this study, we stop the simulation before reaching the Truelove criterion which ensures the isothermal self-gravitating medium is not collapsing. }

{\ch Foreground absorptions or background emissions have little effects in limiting the predicting power of VGT. Even though foreground absorptions can cause significant change in intensity structure of the observed source, distorting the vectors predicted by VGT, such effects can be easily avoided by separating source out in velocity space. Unlike dust polarization data, which is integrated in the broad frequency domain, VGT can be applied to Moment 0 or Moment 1 maps calculated from the channels that only contain the source. By doing the channel selection, VGT has a degree of freedom to separate out source in the line of sight in velocity space.  } 

{\ch There are also some important issues when applying VGT on radio interferometric data and some preprocessing are needed. The first is the angular resolution of interferometric observations will be slightly distorted due to the non circular beam effect. Using elliptical beam to sample the source would cause the data to have position-angle dependent angular resolution. The direction along the major axis of the elliptical beam would have lower angular resolution, causing the structure to be slightly elongated in the direction of the major axis of the elliptical beam. This would cause a systematic error when VGT is applied. In order to obtain unbiased sampling data before applying VGT techniques, it is advised that observers convolve the beam into circular beam before applying VGT to eliminate this systematic error. Also, proper handling of noise in real observational data is required. Observers should perform a careful job to remove all the radio artifacts due to side-lobes, any leftover side-lobe would cause systematic error in VGT. If the deconvolution is done correctly, the noise in the radio data should be random Poisson noise. Random white noise do not produce a systematic error in VGT, however it adds to the flat background when fitting a shifted Gaussian in the block averaging phase. As a criterion, if the constant area is larger than the Gaussian area fitted, then the VGT prediction in that block is discarded. Additional smoothing can be used to averaged out this random noise, however the spatial resolution of VGT would decrease. Observers should choose a smoothing scale to have the right balance between spatial resolution and the accuracy in the Gaussian fitting in block averaging.}

{\ch \subsection{Possible effects of driven and decaying turbulence simulation on the performance on VGT}}

{\ch Supersonic turbulence played an important role in regulating the dynamics of molecular clouds and star formation \citep{1981MNRAS.194..809L}. Currently there are two ways to simulate turbulence in molecular clouds, and each is closely tied to a different view on cloud dynamics. Decaying turbulence simulation initiates a turbulence field and allows it to decay. Without further injection of energy the turbulence will decay with characteristic time scale of roughly 1 crossing time \citep{2004RvMP...76..125M}. After losing the turbulence support, the molecular cloud would undergo global collapse and trigger star formation. To obtain the observed low star formation rate, in this picture the molecular cloud must be a short lived transient structure \citep{2000ApJ...530..277E}. Driven turbulence simulations on the other hand, allows longer-lived molecular clouds and the turbulence energy is constantly injected to the system. Such driven turbulent system has been observed in L1555 low mass star forming region \citep{2008ApJS..174..202S}. }

{\ch In this study, driven turbulent simulations are used to study the behavior of VGT. VGT method is expected to perform better in driven turbulent simulations comparing to the decaying turbulent simulations. This is because after losing turbulent pressure support, the density field of decaying turbulent simulations are expected to be more clustered around the collapse center. When turbulent driving mechanism is turn off, the density profile is expected to evolve closer to a free fall profile $\rho \propto r^{-1.5}$ \citep{2008ApJ...686.1174O}. Stronger gravitational effect is expected in the decaying turbulent simulations resulting the more pronounced distortion of VGT predicted vectors. Denser tracers such as $^{13}$CO and C$^{18}$O are expected to be strongly effected.}

{\ch Driven turbulence simulations on the other hand introduces unphysical driving force, which will resulted in the large artificial mass flux between low density regions and high density regions. High mass flux implies that more materials are effected by turbulent eddies and hence increase the performance of VGT. Driven turbulence simulations are expected to have a steeper velocity profile $\sigma \propto r^{0.5}$ compare to decaying turbulence simulations \citep{1999ApJS..125..161J,2008ApJ...686.1174O}. The steeper slope in velocity profile ensures the magnitude of gradient is large enough to perform VCG. The increase in mass flux also resulted in a flatter density profile $M(R) \propto R$ in driven turbulent simulation \citep{2008ApJ...686.1174O}. Increase in mass flux artificially increases the intensity in the diffuse region. Since Moment 1 is intensity weighted average velocity and Moment 0 is an intensity map, this would result the magnetic field traced along the line of sight being slightly biased towards the diffuse region. Hence improve the performance of $^{12}$CO, a low density tracer, in both IG and VCG method. }

{\ch In both driving and decaying turbulence simulations $^{12}$CO is expected to perform better than $^{13}$CO, C$^{18}$O. In the driving turbulence case, $^{12}$CO is enhanced by the increase in mass flux in the diffuse regions. In the decaying turbulence simulation case, dense tracers such as $^{13}$CO and C$^{18}$O are more strongly affected by the stronger collapsing motion.}

\subsection{Extracting 3D magnetic field structure by utilizing multiple molecular tracer data}

 Molecular tracer maps with different optical depth provide the spectroscopic information of gas dynamics up to certain line-of-sight depth.  As shown in \ref{fig7}, some tracers are optically thick in some velocity channels. In these fully saturated channels, the magnetic field information is not traced.  This motivates us to investigate an important question: whether or not we can use the VGT method to trace different layers of magnetic fields in the line of sight by utilizing multiple molecular tracer data.
 
The concept of "gradient tomography" was first discussed in \cite{LY18b} by considering the effective accumulation line-of-sight deepness of synchrotron polarization data with the different wavelength. Both the synchrotron polarization data with the presence of strong Faraday Rotation effect and the gas spectroscopic data with the presence of optically thick radiative transfer effect share the same concept that the contribution of gas dynamics with line-of-sight deepness larger than some certain physical boundary would be effectively noise. \cite{LY18b} showed that, by stacking the gradient maps from the polarized synchrotron intensities measured from different frequencies, one can create the 3D tomography information of the magnetic field. The number of layers in the gradient tomography completely depends on how many individual frequency measurements one has taken for the synchrotron data. The analogous idea can actually be implemented in the case of multiple molecular tracer maps but in a much coarser content. For example, it is theoretically possible to stack the gradient map from \cotw, \coth, C$^{18}$O to create a 3-layer tomography map which is shown observationally in \cite{MT18}.

One might also question whether the 3D magnetic field is measurable if stacking multiple dust polarization emissions in different wavelength similar to \cite{LY18b}. However, dust polarimetry faces several limitations in tracing magnetic field. Not only the dust grain fails to align at high optical depth if there are no illuminating sources inside the cloud \citep{2007JQSRT.106..225L}, but also the far infrared polarization suffers from the confusion effect when the cloud is in the galactic plane. It is therefore not practical to stack multiple dust emission maps with different wavelength in acquiring the 3D magnetic fields similar to the 3D synchrotron polarization gradients \citep{LY18a} or the Faraday tomography method \citep{1966MNRAS.133...67B}. 

In any case, the 3D studies of absorbing species are complementary to the studies in HI and synchrotron polarization. This gives the 3D structure of magnetic field in the galactic magnetic ecosystem.

\subsection{Synergy with the latest development of VGT}

\subsubsection{Improving the accuracy of B-field orientation tracing by PCA-VGT}

In terms of tracing the magnetic field in ISM, the accuracy is the most important aspect. \cite{2018arXiv180208772H} demonstrated that the accuracy of VGT can be significantly improved by the Principal Component Analysis (PCA), which is widely used in image processing. In this work, the VGT also shows a robust performance in tracing the magnetic field with the presence of molecular media. Furthermore, we expect that the PCA can also extract the spectroscopic information which is most valuable for VGT considering the emission from molecular media and then the synergy of VGT and PCA would further improve the accuracy.

\subsubsection{Acquisition of $M_s$, $M_A$}

The magnetization of the interstellar medium is also one crucial aspect of the star formation theory. Recently, the VGT has been introduced in \cite{2018arXiv180202984L} to obtain a reliable estimation of the magnetization of the media in HI data. In this work, we show that different molecular tracers contain individual information due to their own optical depth. Hence, we see the possibility to construct the 3D strength map of the magnetic field by combining different molecular tracers.

\subsubsection{The relationship between effective optical depth $\tau$ and channel optical depth and its implications on VGT}

\begin{figure}[h]
\centering
\includegraphics[width=1.0\linewidth,height=.7\linewidth]{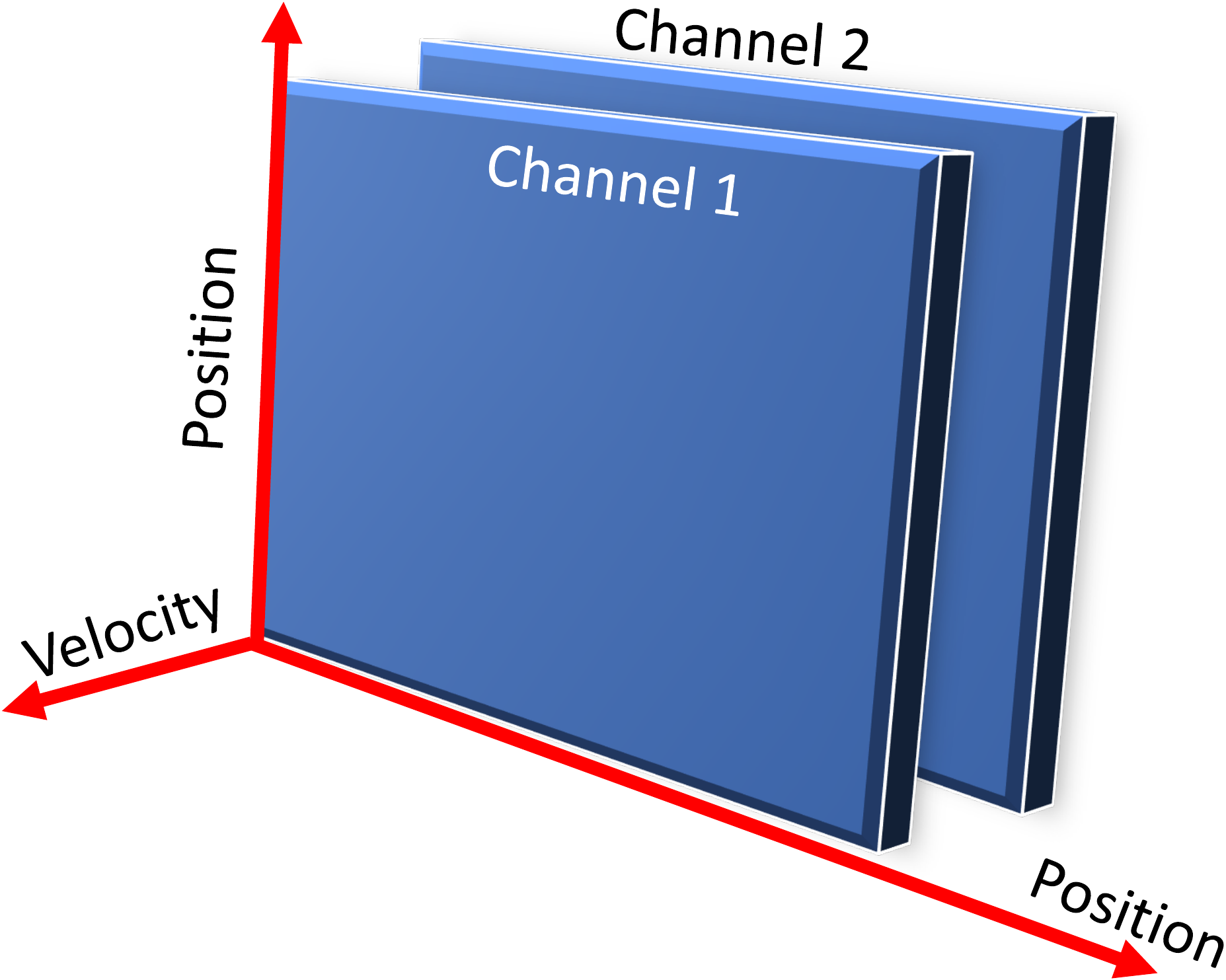}
\caption{A cartoon diagram of a 2 channel PPV cube.
}
\label{fig8}
\end{figure}

In \autoref{fig7}, we plot the averaged intensity in each velocity channel, and this spectrum offers great insight on how IG operates. It tells us the contribution of the magnetic fields at each channel traced by IG method. It also explains why $^{12}$CO performs better than $^{13}$CO and C$^{18}$O. 

To illustrate the concept, we first demonstrate the relationship between effective optical depth $\tau_{eff}$ for moment 0 map and optical depth in individual velocity channels. Consider a simple two channel PPV cube shown in \autoref{fig8}. In the PPV cube the temperature is set as uniform 10K. When we add the two channels to calculate the moment 0 map, the effective optical depth added as following:

\begin{equation}
\begin{aligned}
\frac{I_1+I_2}{2} &=I_{bg}e^{-\tau_\nu} +\frac{I_{10K} [1-e^{-\tau_1}]+I_{10K} [1-e^{-\tau_2}]}{2},\\
I_{avg} &=I_{bg}e^{-\tau_\nu} +I_{10K}-\frac{I_{10K} [e^{-\tau_1}+e^{-\tau_2}]}{2},\\
e^{-\tau_{avg}} &=\frac{[e^{-\tau_1}+e^{-\tau_2}]}{2}.\\
\end{aligned}
\label{eq4}
\end{equation}

In the case of Channel 1 is much more optically thicker than Channel 2 ($\tau_1 \gg \tau_2$), $e^{-\tau_{avg}}\approx \frac{e^{-\tau_2}}{2}$ or roughly $\tau_{avg} \approx \tau_{2}$. This simple example illustrates an import implication on IG: Optically thick channels contributed less in the effective optical depth calculations. {\bf In the IG method, most of the magnetic fields traced is in the tail distribution of \autoref{fig7}.} In the central saturated region little or not magnetic fields are traced and this resulted $^{12}$CO mainly traced magnetic fields at hight velocity diffuse region and have better IG performance. The more in-depth discussion of saturation and which region of magnetic field is traced will be addressed in the future paper (C.-h. Hsieh et. al in prep). 

\subsubsection{Locating the self-gravitating regions}

Furthermore, since VGT is available even with the presence of self-gravity, it is possible to understand how magnetic fields behave inside the Giant Molecular Clouds (GMCs) in the stage of collapsing and help us to shape the star formation theory into a better form. On the other hand, the spectroscopic data is obtained by using molecular tracers, such as $^{12}$CO, $^{13}$CO, C$^{18}$O and so on, which contains individual information from low-density regions to high-density regions. By using VGT, it is possible to extract the magnetic field morphology in different regions or layers and then construct the 3D magnetic field morphology with multi-stages. One obstacle is that the velocity field would be distorted in a collapsing
region. Hence, in collapsing regions, we usually need to apply a compensation of the distortion to VGT for accurate tracing. 

\subsection{Comparison with earlier works}

The VGT technique has been initially introduced and studied without taking into account the effects of self-absorption of radiation \citep{YL17b,YL17a}. Such studies are most relevant to diffuse HI and other media with low absorption. {\ch For molecular gas tracers such as CO used in this paper, radiative transfer is crucial in order to get the correct result for VGT.}
\cite{2017arXiv170303035G} estimated the ability of VCGs in tracing magnetic field with the presence of absorbing media for the case of $^{13}$CO J=2-1 emission. They also demonstrated that the VGT is able to trace the magnetic field in media with different CO abundances, densities and optical depths. However, the change of CO abundances and densities might introduce the non-LTE effects {\ch and saturation effects}. In this work, we {\ch extended \cite{2017arXiv170303035G}'s study with typical CO abundances inferred based on observation \citep{1999RPPh...62..143W,ToolsofRadio_astronomy,1999RPPh...62..143W}. We} study both the IGs and VCGs and we also explore the effect from different {\ch molecules} $^{12}$CO, $^{13}$CO, and C$^{18}$O with emission line J=1-0. 

% {\ch in their work they multiply and divide the density and CO abundances by factor of 30 to generate different cases. This approach is not ideal when applying to observational result. The CO abundances is constrained by observation unless other chemical or physical mechanisms are proposed in the GMC \citep{1999RPPh...62..143W,ToolsofRadio_astronomy,1999RPPh...62..143W}.}

The optical depths are different for different species. In term of the applicability of multi-tracers in observation, we numerically demonstrate the possibility to obtain 3D magnetic field morphology by combining different molecular tracers. Furthermore, we consider the effect of weak self-gravity and show that VGT is still applicable. The effect of line-saturation effects in the VGT will be investigated in our future work.

{\ch The Histogram of Relative Orientation (HRO) is a statistical way to determine the relative orientation of magnetic field and density gradients \citep{Soler2013}. To avoid confusion, one should distinguish the IGs and HRO that IGs is the technique to trace magnetic fields in space, while HRO does not have this ability. IGs can even be use to study shocks and regions of gravitational collapse, etc. \citep{YL17b,LY18a}, but does not require any polarization data to get this information. HRO, on the contrary, compares the relative orientation of the density gradients and the polarization directions as a function of column density. Since HRO is only a tool to study the statistical correlation , we expect to get more information  between the VCGs and the magnetic field by using HRO.
}

\section{Conclusion}
\label{sec:7}

In this work, we estimate the ability of VGT in tracing the magnetic field with the presence of molecular medium ($^{12}$CO, $^{13}$CO, and C$^{18}$O), as well as with and without self-gravity cases. We demonstrate that the VGT shows robust performances in the presence of the radiation transfer and self-gravity. As a part of the VGT we use IGs, which are a derivative approach based on applying the VGT procedures to emission intensities. To summarize:
\begin{enumerate}

\item VGT method is the most accurate in $^{12}$CO. For $^{13}$CO, C$^{18}$O the dispersion of the relative angle is larger as shown in \autoref{tab:sim}. This is because $^{12}$CO only traces optically thin channels which corresponds to the diffuse region. 

\item Centroid gradient method works better than the intensity gradient method. Velocity weights on high-velocity channels increase the weighting for the diffuse region.

\item As Sonic Mach number $M_S$ increases, the magnetic field direction get more perpendicular to the intensity gradient. This effect to less {\ch extent} is also present for the centroid gradient.

\item For systems that have values within $0.02< \beta < 0.2$ and $M_S \geq 1.0$ the VGT method has uncertainty less than 20$^o$ under 1 $\sigma$ (68.27$\%$) confidence interval and can make very accurate B field predictions.

\item For density $n(H_2)(cm^{-3})$ between $0.004\sim70$ cm$^{-3}$, applying VGT methods to self-gravity {\ch data} shows slightly larger dispersion compare to the data without self-gravity.

\end{enumerate}

\noindent
{\bf Acknowledgment.} {\ch CHH and SPL acknowledge support from the Ministry of Science and Technology of Taiwan
with Grant MOST 106-2119-M-007-021-MY3. AL acknowledges the support of the NSF AST 1816234 and the NASA AAG1967 grants.}

\appendix
\section{SPARX}

SPARX stands for "Simulation Package for Astronomical Radiative Transfer (Xfer)". As part of the CHARMS (Coordinated Hydrodynamic and Astrophysical Research, Modeling, and Synthesis) initiative which focuses at bridging theory and numerical astrophysics with observational astrophysics at the 
Theoretical Institute for Advanced Research in Astrophysics (TIARA) in ASIAA, SPARX is a multi-purpose radiative transfer calculation tool. It is designed to generate synthetic (non-)LTE atomic and/or molecular spectral and dust continuum images.

SPARX solves the specific intensity $I_{\nu}$ at a given frequency $\nu$ with the standard radiative transfer equation,

\begin{align}
\frac{dI_{\nu}}{ds}=-\kappa_{\nu} I_{\nu} + \epsilon_{\nu}
\end{align}

where $\kappa_{\nu}$ is the absorption coefficient and $\epsilon_{\nu}$ is the emission coefficient at the given frequency $\nu$. The above equation can be rearranged into the following form:

\begin{align}
\frac{dI_{\nu}}{d\tau_{\nu}}=-I_{\nu} + S_{\nu}
\end{align}

where $d\tau_{\nu} \equiv \kappa_{\nu} ds$ is the optical depth, and $S_{\nu} \equiv \epsilon_{\nu}/\kappa_{\nu}$ is the source function.
The intensity can be evaluated numerically by

\begin{align}
I_{\nu}= \sum_{ \forall cell} S_{\nu} (1-e^{-\Delta\tau})e^{-\tau}
\end{align}

We note that $\kappa_{\nu}$ and $\epsilon_{\nu}$ are related to Einstein A and B coefficients and the gas density $n$:
 \begin{align}
 \kappa_{ij}(\nu)  = n_iA_{ij}\phi(\nu)
 \\
 \epsilon_{ij}(\nu) = (n_jB_{ji}-n_iB_{ij})\phi(\nu)
 \\
 S_{ij} = \frac{ n_iA_{ij}}{n_jB_{ji}-n_iB_{ij}}
\end{align}
  where i and j denote the starting and ending energy states of the molecular transition under consideration. 
$n_i$ is the gas density at energy state $i$, and $\phi(\nu)$ is the Doppler broadening function:
\begin{equation}
\begin{aligned}
\phi(\nu) &= \frac{c}{b\nu_{0}\sqrt[]{\pi}}\exp\left(\frac{-c^2(\nu-\nu_0)^2}{\nu_0^2 \sigma^2}\right)\\
\end{aligned}
\label{eqA3}
\end{equation}
in which $\sigma$ is the line-width summed by the thermal speed and the turbulent speed and $c$ is the speed of light. 

The Einstein coefficients themselves in the above equations are related in the following manner,
\begin{align}
\frac{g_j B_{ji}}{g_iB_{ij}}  = 1
\\
\frac{A_{ii}}{B_{ij}} = \frac{ 8\pi h{\nu}_{0}^3}{c^3}
\end{align}
where $g_i$ is the statistical weight of the energy state i, $h$ is the Planck constant. In this work , we {\ch consider} the rotational transitions of carbon monoxide which have statistical weight of $g_J = 2J+1$. Additional molecular data required for the calculation are retrieved from the LAMDA database (http://home.strw.leidenuniv.nl/~moldata/).

The level populations $n_i$ and $n_j$ of energy states $i$ and $j$ required for evaluating $\kappa_{\nu}$ and $\epsilon_{\nu}$ in the radiative transfer equations should be solved through detailed balancing. This, in term, depends on the incoming radiation. Therefore, in the general non-LTE calculation, the specific intensity ($I_{\nu}$) and the mean radiation field (integral of $I_{\nu}$) and the level populations $n_i$ are solved iteratively.

In this work, the molecular transitions under consideration should meet the LTE assumption. The molecular level populations can therefore be described by the Boltzmann distribution: 
\begin{align}
\frac{n_j}{n_i} = \frac{g_j}{g_i} e^{-\frac{-h{\nu}_0}{kT}}
\end{align}
or
\begin{align}
\frac{n_{i}}{N} &= \frac{g_{i}e^{\frac{-E_{i}}{kT}}}{Z}, 
\end{align} 
where $N$ is the total molecular density and $Z$, the partition function, can be expressed as 
\begin{align}
\quad Z &= \sum_{i \in \textit{ensemble}} g_{i}e^{\frac{-E_{i}}{kT}}\\
\end{align}

For further detailed description of the SPARX software and the benchmark problem (Zadelhoff et al. 2002) the package has been tested against, we refer to the software website (https://sparx.tiara.sinica.edu.tw/).

\newpage
\section{$^{13}$CO cumulative plots}
\begin{figure}[tbh!]
\centering
\includegraphics[scale=0.275]{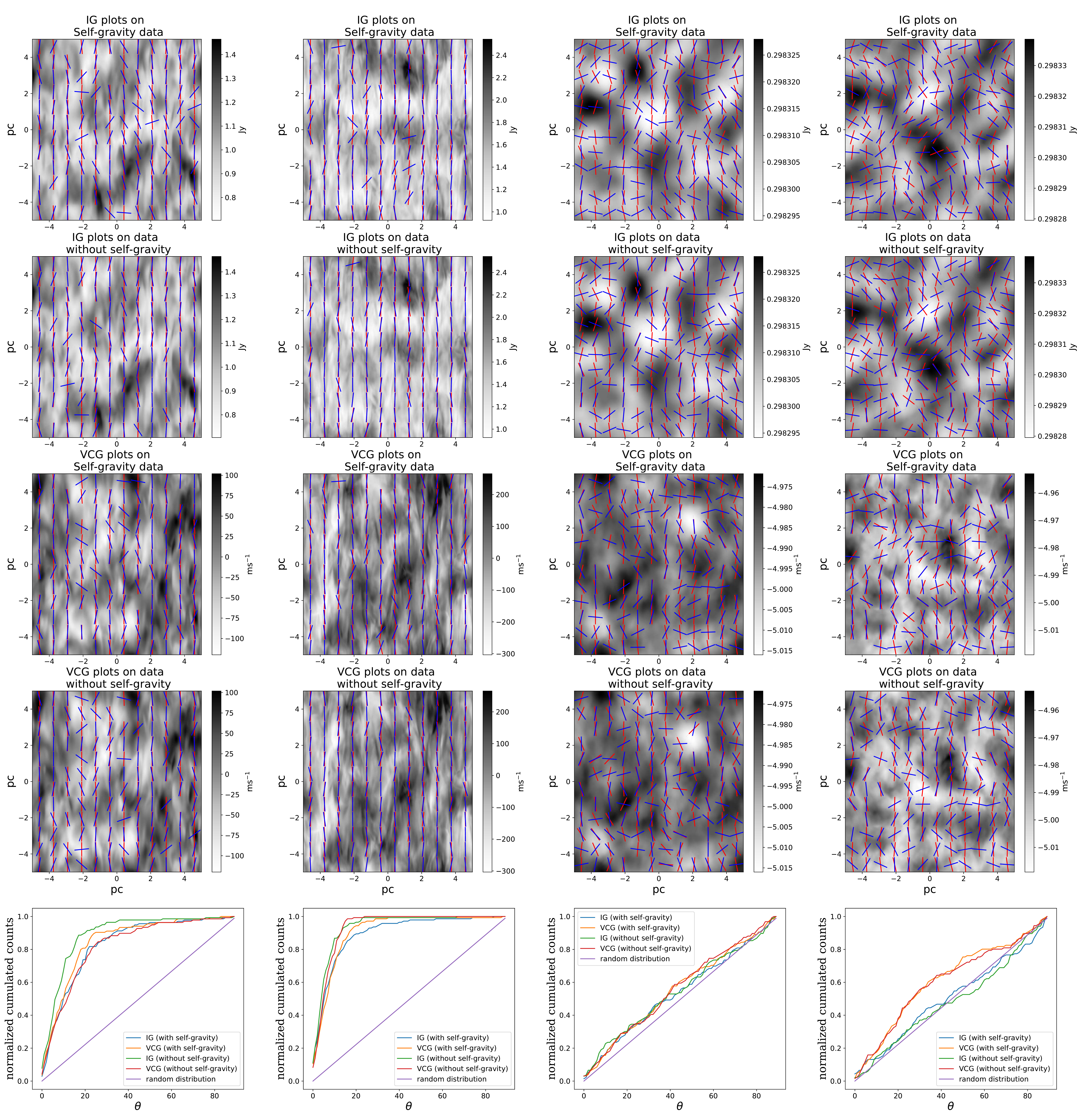}
\caption{From left to right: b12 ($M_S=0.92$, $M_A=0.09$), b13 ($M_S=1.95$, $M_A=0.18$), b51 ($M_S=0.05$, $M_A=0.52$), b52 ($M_S=0.10$, $M_A=1.08$) $^{13}$CO data set. For each method (IGs and VCGs) and each data (with or without self-gravity) we plot the 2D vector plots and the statistical results. The blue vectors represent the projected B fields from the simulation, and the red vectors represent the VGT predicted B field direction. The relative angle between the simulated B field and the VGT predicted direction is shown in the normalized cumulative plots.}
\end{figure}

\section{C$^{18}$O cumulative plots}

\begin{figure}[tbh!]
\centering
\includegraphics[scale=0.275]{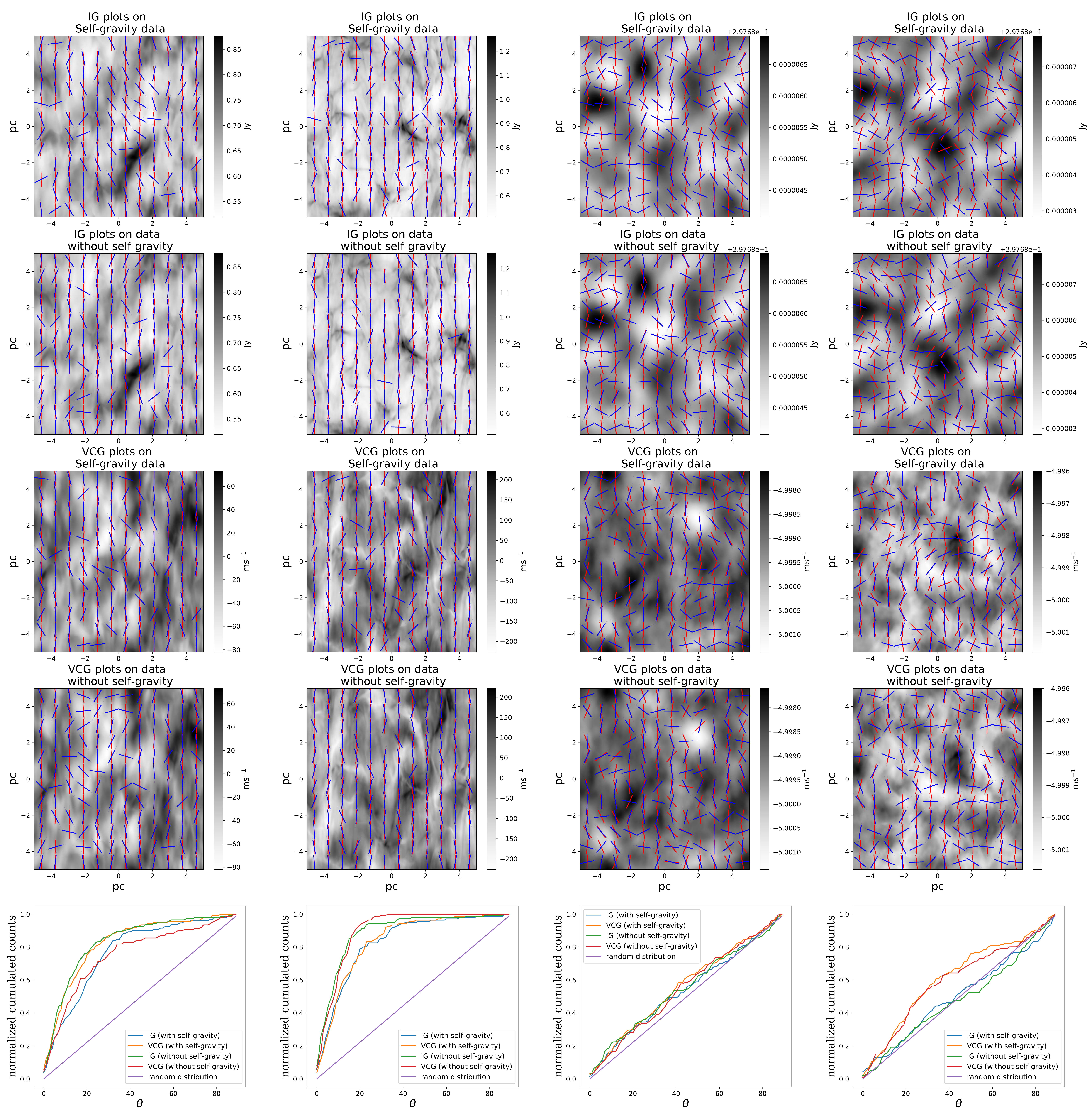}
\caption{From left to right: b12 ($M_S=0.92$, $M_A=0.09$), b13 ($M_S=1.95$, $M_A=0.18$), b51 ($M_S=0.05$, $M_A=0.52$), b52 ($M_S=0.10$, $M_A=1.08$) C$^{18}$O data set. For each method (IGs and VCGs) and each data (with or without self-gravity) we plot the 2D vector plots and the statistical results. The blue vectors represent the projected B fields from the simulation, and the red vectors represent the VGT predicted B field direction. The relative angle between the simulated B field and the VGT predicted direction is shown in the normalized cumulative plots.}
\end{figure}

\bibliographystyle{apj.bst}

\end{document}